\RequirePackage{fix-cm}
\documentclass[smallcondensed]{svjour3}     
\smartqed  
\usepackage{graphicx}
\usepackage{subfigure}
\usepackage{algorithm}
\usepackage{algpseudocode}
\usepackage{amssymb}
\usepackage{amsmath}
\usepackage{bbding}

\usepackage{hyperref}
\hypersetup{
	colorlinks={blue},
	linkcolor=blue
}
\begin{document}

\title{Enhancing Stock Market Prediction with Extended Coupled Hidden Markov Model over Multi-Sourced Data
}


\author{Xi Zhang$^1$         \and
        Yixuan Li$^1$ \and
        Senzhang Wang$^2$ \and
        Binxing Fang$^{1,}$$^3$ \and
        Philip S. Yu$^{4,}$$^5$
}


\institute{\Envelope \ \ \ Xi Zhang\\
\email{zhangx@bupt.edu.cn} \\
              $^1$ \ \ Key Laboratory of Trustworthy Distributed Computing and Service (Beijing University of Posts and Telecommunications), Ministry of Education, Beijing, China. \\%
              $^2$ \ \ Department of Computing, The Hong Kong Polytechnic University, Kowloon, Hong Kong  \\     
              $^3$ \ \ Institute of Electronic and Information Engineering of UESTC in Guangdong, China. \\
              $^4$ \ \ Department of Computer Science, University of Illinois at Chicago, IL, USA.\\
              $^5$ \ \  Institute for Data Science, Tsinghua University, Beijing, China.         
}

\date{Received: date / Accepted: date}

\maketitle

\begin{abstract}
Traditional stock market prediction methods commonly only utilize the historical trading data, ignoring the fact that stock market fluctuations can be impacted by various other information sources such as stock related events. Although some recent works propose event-driven prediction approaches by considering the event data, how to leverage the joint impacts of multiple data sources still remains an open research problem. In this work, we study how to explore multiple data sources to improve the performance of the stock prediction. We introduce an Extended Coupled Hidden Markov Model incorporating the news events with the historical trading data. To address the data sparsity issue of news events for each single stock, we further study the fluctuation correlations between the stocks and incorporate the correlations into the model to facilitate the prediction task. Evaluations on China A-share market data in 2016 show the superior performance of our model against previous methods.
\end{abstract}

\section{Introduction}
\label{intro}

The capability of predicting the stock price movement directions can offer enormous arbitrage profit opportunities and thus attract much attention from both academia and industry. Conventional quantitative trading prediction methods are mostly based on the historical trading data such as prices and volumes. According to the Efficient Market Hypothesis (EMH)~\cite{fama1965behavior}, stock prices are the reflection of all known information. Therefore, merely relying on historical quantitative data are not sufficient for an accurate prediction. As more and more investors obtain information from social media~\cite{MMRate,ZHANG2017}, the indicators obtained from Web news articles and social networks can also have significant impacts on the stock prices, and thus such factors that can derive the stock price fluctuations must be considered. As such, there are growing research interests in exploring financial text documents such as news articles, financial standings to facilitate the stock prediction task.

Previous studies showed that news events, such as corporation acquisition and earning announcement, could have significant impacts on the stock prices~\cite{cutler1988moves,tetlock2008more,luss2015predicting,xie2013semantic,wang2014semiparametric,peng2015leverage}. Recent advances in NLP techniques enable the capability in extracting events from news articles, which motivates event-driven stock prediction approaches~\cite{das2007yahoo,tetlock2007giving,si2013exploiting}. However, these prediction capabilities are limited by the following factors. First, the events may only account for a small part of stock volatility, making it insufficient for estimating stock market alone. Second, the events extracted from the news are usually quite sparse, resulting in unreliable and unsustainable predictive power. Third, it is challenging to extract and represent the events from various types of free texts~\cite{lavrenko2000mining,kogan2009predicting,schumaker2009textual}. For example, the same event may be described in different ways by different news articles and thus is prone to be identified as different events. Thus, replying on events alone for stock prediction may not be sustainable, and how to effectively extract and represent the event information from free texts is still under exploration.


Due to the complexity of the stock market, using a single data source is insufficient to fully capture the multiple factors that affect stock fluctuations. Thus it is natural to investigate how to combine the historical qualitative data with various events to perform a better prediction. It is challenging because the distinct characteristics including the data formats and temporal granularities corresponding to different data sources. Nowadays, information fusion approaches have been proposed in various domains such as urban computing~\cite{wang2016enhancing} and cybersecurity~\cite{alsmadi2016information}, but their application in stock market prediction by incorporating web data is rarely covered. In addition, as the stock-related data is commonly served with time series models, how to effectively fuse multi-source information regarding time series makes the problem even more difficult. Although some recent works propose tensor-based models to combine events, quantitative data, and sentiments~\cite{li2015tensor}, they use some simple event features that cannot tackle the event sparsity problem, and thus largely limit the power of the prediction model.

To address the aforementioned challenges, in this paper, we integrate the events extracted from news articles together with historical quantitative stock price data to enhance stock market prediction. Motivated by the successful applications of Hidden Markov Models (HMM) in various time sequential scenarios, in this work, we propose a novel Extended Coupled Hidden Markov Model (ECHMM) to effectively fuse the two types of data for stock prediction. Specifically, in contrast to traditional HMM where each hidden state is associated with only one observation, each hidden state in ECHMM is associated with two different types of observations. That is, the stock price and stock related event are fused in one unified framework. In addition, this framework also incorporates the correlations between stocks, which are intuitively important but largely neglected by previous works. We consider the current price state of each stock depends on not only its previous price state but also the previous price states of its correlated stocks (connected neighbor nodes in ECHMM). With the correlations between stock prices, instead of treating each stock prediction task independently, we propose to predict the future price states of multiple correlated stocks simultaneously, which can potentially alleviate the sparsity of events and provide better predictive performance.

The main contributions of this work can be summarized as follows:
\begin{itemize}
\item[1)] To fully leverage the data from multiple sources, we propose a stock prediction framework based on HMM model by integrating heterogeneous information including Web news and historical quantitative data.
\item[2)]  To alleviate the event sparsity problem, we incorporate stock correlation information in this framework by sharing knowledge among multiple stocks.
\item[3)] We evaluate our framework on the China A-share market dataset, and the results show that the prediction performance can be significantly improved by integrating the event and stock correlation information.

\end{itemize}

The remaining of the paper is organized as follows. Section~\ref{relatedwork} introduces the related work. In Section~\ref{preliminary}, we give the preliminary and problem definition. Section~\ref{framework3} describes the system framework. We elaborate the extended Coupled Hidden Markov Model and how to make predictions by our model in Section~\ref{model}. In Section~\ref{experiment}, we show the effectiveness of the proposed approach by evaluating on real data. Finally, we conclude the paper in Section~\ref{conclusion}.

\section{Related work}
\label{relatedwork}
Most of previous studies utilize historical time-series prices to predict the future prices of instruments in financial market and make predictions with various models~\cite{patra2009computationally,saad1998comparative,roman1996backpropagation,chen2015lstm,jia2016investigation,Chiang2016,Chong2017,Goumatianos2017}. An intelligent decision support system combines influence digram generator, probability assessor, value function generator to help decision-makers make better investment decisions~\cite{Poh2000}. Recent works begin to explore other data sources to improve the predictive power. Weng et al.~\cite{Weng2017} propose to integrate online data sources including Google news counts and Wikipedia traffic with traditional technical indicators to make predictions. 

There are a line of works using event-driven models to make stock predictions, which commonly extract events from news articles or news titles. Hogenboom et al.~\cite{hogenboom2011overview} give an overview of event extraction methods. Akita et al.~\cite{akita2016deep} use Paragraph Vector to convert newspaper articles into distributed representations and apply LSTM to model the temporal effects of past events on opening prices of stocks in Tokyo Stock Exchange. Nguyen et al.~\cite{nguyen2013event} formulate a temporal sentiment index function to extract significant events and then analyzed the corresponding blog posts using topic modeling to understand the contents. Ding et al.~\cite{ding2014using} applied the Open IE tool to extract structured events from texts, and used the off-the-shelf classification algorithms for prediction. Ding et al.~\cite{ding2015deep} trained event embeddings with a neural tensor network and then used a deep convolutional neural network (CNN) to model influences of events with various temporal granularities. A deep neural model is proposed to understand an events economic value by measuring the content of financial news~\cite{changmeasuring}. Temporal properties of news events that have short-term and long-term influences on stock prices are modeled in~\cite{yoshihara2014predicting}. However, these studies didn't consider the correlations among stocks. Correlated stocks commonly share similar impacts of a news event, and exhibit co-movement in prices, which can help to make a better prediction. The company correlations are learned through a consensus of correlations based on multiple representations extracted from Twitter in~\cite{Zhang2014Discovering}, but they have a different focus, that is, they didn't utilize the correlations to predict the stock market.


There are also a few studies that analyze the impacts of sentiments on stock market volatility. One main data source for sentiment analysis is the news articles~\cite{schumaker2009textual,feldman2011stock,schumaker2009quantitative,li2014news}, and the other common data source is the social media~\cite{si2013exploiting,Oliveira2017,nguyen2015topic,zhang2011predicting}. Oliveira et al.~\cite{Oliveira2017} use sentiment and attention indicators extracted from microblogs and survey indices to predict stock market behavior. Topics and related sentiments are extracted from the texts in a message board, which are provided to facilitate stock prediction~\cite{Hai2015}. 
With Twitter data, the social relations between stocks are exploited to build a stock network based on the co-occurring relationships~\cite{si2014exploiting}. Then a lexicon-based sentiment analysis method is applied to compute the sentiment score for each node and each edge. The sentiment time series and price time series are used for prediction. However, these studies didn't take the impacts of the news events into account.

In addition to the aforementioned studies that consider either news events or sentiments separately, recent work begin to model their joint impacts. Events and sentiments are integrated in a tensor framework in~\cite{li2015tensor}, but they didn't utilize the correlations between stocks. To address this issue, a coupled matrix and tensor model is proposed in~\cite{Zhang2018Improving}, which predicts the movements of multiple stocks simultaneously through their correlations. However, it didn't model the time series information. Both~\cite{li2015tensor} and~\cite{Zhang2018Improving} are compared with our proposal as baselines.

\underline{Stock market prediction is highly related to time series models such as Hidden} \underline{Markov models (HMMs)~\cite{rabiner1989tutorial,abushariah2010natural,elmezain2009hidden,nickel2011using} and deep LSTM networks~\cite{Zhu:2016:CFL:3016387.3016423}.} Hassan and Nath~\cite{hassan2005stock} use HMM to predict the closing price on the next day of airline stocks. Gupta et al.~\cite{gupta2012stock} present the Maximum a Posteriori HMM approach to forecast stock values for the next day given historical data. Park et al.~\cite{park2009forecasting} forecast change direction (up or down) of next day's closing price of financial time series using the continuous HMM. Traditional Hidden Markov model just uses a single Markov chain without considering the interactions between different objects, which may lose some useful information. To solve this problem, the coupled hidden Markov model (CHMM) was proposed. Abdelaziz et al.~\cite{abdelaziz2015learning} used CHMM as a state based decision fusion model, which allows asynchrony on the state level while preserving the natural dependency between the audio and video modality. Nefian et al.~\cite{nefian2007coupled} thought coupled HMM was a generalization of the HMM suitable for a large scale of multimedia applications that integrate two or more streams of data. Kumar et al.~\cite{kumar2017coupled} proposed CHMM which provided interaction in state space instead of observation states as used in classical HMM that fails to model correlations between inter-modal dependencies. Wang et al.~\cite{wang2016enhancing} used CHMM to integrate GPS probe readings and traffic related tweets to accurately estimate traffic conditions of an arterial network.


\label{}

\section{Preliminary}
\label{preliminary}

In this section, we first start with some definitions and then make some basic assumptions to facilitate introducing the model.
\subsection{Definitions}

The definitions are given as follows.

\emph{\textbf{Definition 1 A news observation of stock event \emph{$e_{t, s, i}$}.}} We represent the \emph{i-th} news observation of a stock \emph{s} at time \emph{t} as such a tuple \emph{$e_{t, s, i}$ = (f, s, t)}, where \emph{f} is the stock news category, \emph{s} represents the stock ID, and \emph{t} denotes the time of the stock event. 

 \emph{\textbf{Definition 2 A historical stock price observation \emph{$m_{t, s, i}$}.}} We represent the \emph{i-th} historical price observation on stock \emph{s} at time \emph{t} as such a vector \emph{$m_{t, s, i}$ = (p, s, t)}, where \emph{p} is the stock price, \emph{s} is the specific stock and \emph{t} denotes the time.

 \emph{\textbf{Definition 3 A set of correlated stocks \emph{$C_{s}$}.}} Two stocks \emph{$s_{1}$} and \emph{$s_{2}$} are considered as two correlated stocks if the similarity between them is above a predetermined threshold. The stock similarity is calculated based on their co-evolving prices (explain in detail later). For each stock \emph{s}, we calculate its similarity with every other stock, and all the stocks whose similarities with \emph{s} above the threshold will be considered as the correlated stocks with \emph{s}, which form a set of correlated stocks, termed as \emph{$C_{s}$}. Please note that, a stock is considered as a correlated stock with itself.

Please note that the news category in \emph{Definition 1} means a set of news events on similar topics, e.g., on politics or sports. How to obtain the news category (i.e. event class) will be described later in the event extraction section.

\subsection{Basic assumptions}

HMMs can describe the time-series behaviors and have been used extensively in a wide range of applications such as speech recognition and stock market prediction. HMMs are based on a set of unobserved underlying states amongst which transitions can occur and each state is associated with a set of possible observations. Next, we will make some assumptions based on HMM for computational tractability.

\emph{\textbf{Assumption 1 Binary stock states.}} In each time interval \emph{t}, the price movement direction of stock \emph{s} can be represented by a binary value \emph{h} which has two optional values, 1 (rise) or 0 (fall).


\emph{\textbf{Assumption 2 Conditional independence of state transitions.}} Conditioned on the states of stock \emph{s} and the states of its correlated stocks in time interval \emph{t}, the state of stock \emph{s} at time \emph{t+1} is independent from all other current stock states, all earlier stock states, and all past observations.

 \emph{\textbf{Assumption 3 Conditional independence of stock price and stock related events.}} Conditioned on the state \emph{h} of a stock \emph{s}, the stock price on stock \emph{s} is independent from the stock news event occurring on stock \emph{s}.

The second assumption indicates that the state of a stock is only related to its correlated stock states in the last time interval, but independent of the states of the other stocks. The third assumption shows that the two types of observations, stock price and stock related event, are independent to each other and only determined by the current hidden state of the stock.

In this paper, our problem is how to explore the stock correlation information to further improve the performance, and the intuition behind is that the price fluctuations of the stocks can be correlated and the price fluctuation of one stock can affect the price of some other stocks. So our goal is to propose a framework to accurately predict the stock market by fusing the quantitative trading information and news events.

\section{The Framework}
\label{framework3}

\begin{figure}
  \centerline{\includegraphics[width=0.9\textwidth]{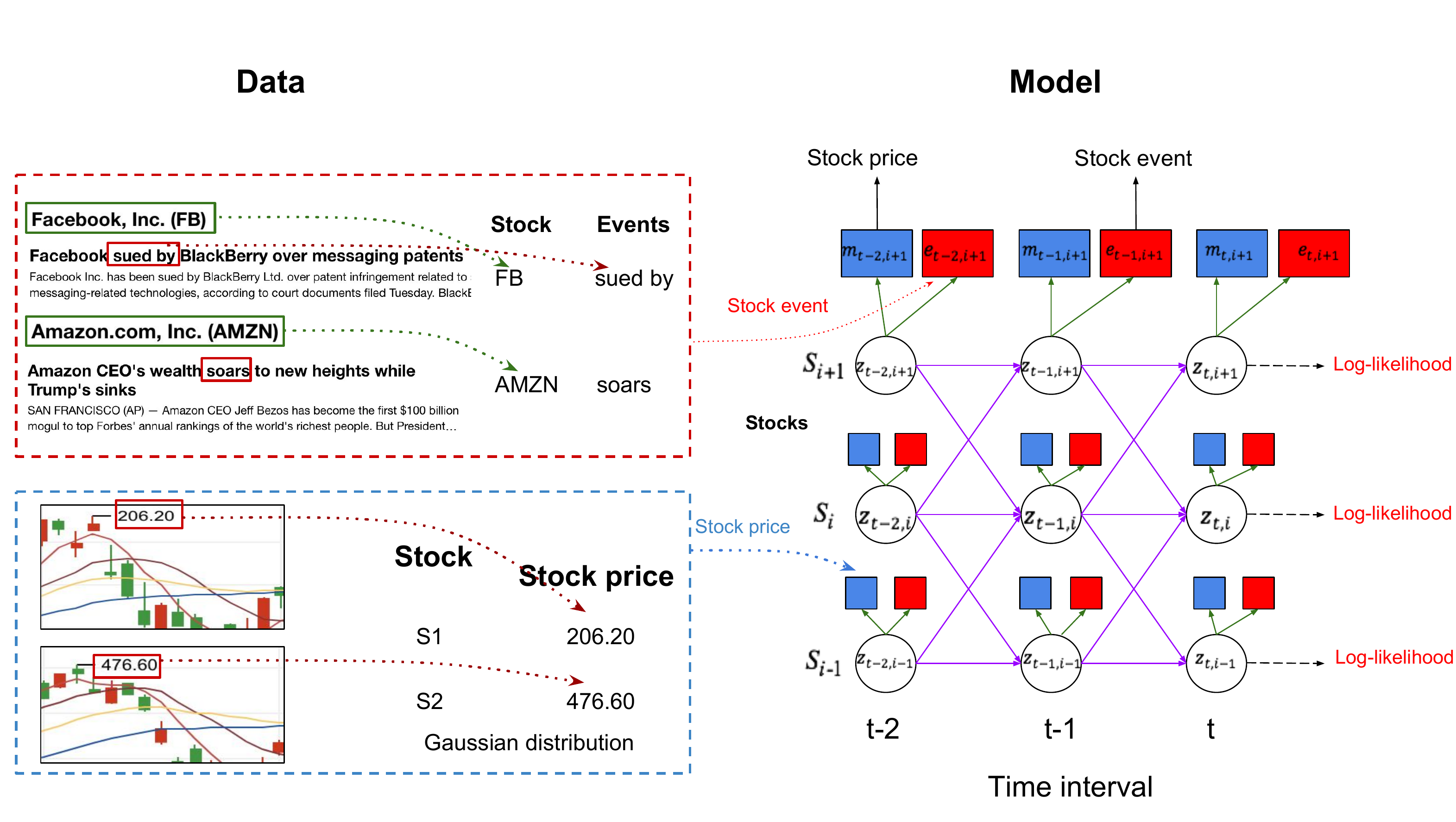}}
  \caption{The framework}
  \label{framework}
\end{figure}

Our proposed framework is shown in Fig.~\ref{framework}, which consists of three parts. The first part is the event extraction and representation part, the second part is the stock price processing, and the third part is the Extended Coupled Hidden Markov Model (ECHMM). We model the temporal conditional dependency of each stock's fluctuations as a Markov process, where the hidden state (circle shown in Fig.~\ref{framework}) indicates the stock price would rise or fall. Each hidden state is associated with two types of observations, namely, stock event observation and stock price observation. In each trading day, we extract the price information from quantitative trading data and event information from news articles and map them to the corresponding stock as the price observation and event observation respectively. Different from traditional HMM model, ECHMM models multiple interaction processes in a unified framework. Thus, each hidden state may depend not only on the previous state of its own but also on the previous states in its neighboring stocks. In our case, neighboring stocks are those correlated stocks. For instance, as shown in Fig.~\ref{framework}, $s_i$ is correlated with $s_{i+1}$ and $s_{i-1}$, and thus the state $Z_{t-1,i}$ is dependent not only on $Z_{t-2,i}$, but also on $Z_{t-2,i-1}$ and $Z_{t-2,i+1}$. If a stock has no correlated stocks, its state would only depend on the previous state of its own. We then explain how to obtain the stock correlation information in the following part.
%
%

\emph{\textbf{Stock correlation.}} Correlation among stocks can be defined in different ways. In this work, we simply identify multiple stocks that co-evolve in prices as correlated stocks. Specifically, we calculate two stocks' similarity based on their p-change values. The p-change value of a stock in a trading day is defined as the change rate between its closing prices in the current day and in the previous day. By concatenating each stock's p-change for each trading day during a period, we can obtain a p-change curve for each stock. Then correlation is obtained by applying Pearson Correlation Coefficient for each pair of stocks on their p-change curves, reflecting the co-evolving movements not only considering the fluctuation direction but also the fluctuation range. If the Pearson Correlation Coefficient is equal to or above a predetermined threshold,  the corresponding two stocks would be identified as correlated ones, and their chains would interact with each other in the ECHMM framework.

\emph{\textbf{Event extraction.}} We then introduce how to extract and represent the stock related events. Commonly, the titles of the news articles in most cases can cover the essential information in the news~\cite{ding2014using}, and thus we extract the events from the titles of the news. For the sake of simplicity, we only use the verb and gerund in the news titles to represent an event as they are quite representative and informative. For example, in the news title ``Microsoft to acquire LinkedIn", the verb "acquire" can denote the event quite well. The subject and object in the titles are omitted in this work as each piece of the news has already been assigned to the specific stock by our data provider. Our news data source will be described in detail in the experimental section.

To extract events, we first use Jieba~\footnote{https://github.com/fxsjy/jieba}, an open-source Python component for Chinese text segmentation, to segment the news titles to obtain the verb and gerund. We then train the word embeddings on those verbs and gerunds with word2vector~\cite{mikolov2013distributed} by using Chinese finance news corpus~\cite{Efficiency2016Effective}, and set the dimension size as 100. If a stock has more than one news in one day, there will be more than one word embeddings for a stock in a day and thus we will average multiple word embeddings, ensuring that there is at most one word embedding for each stock in each day. After that, we apply the k-means method to cluster the embeddings and obtain 300 clusters. Thus, if a stock has one or more events in one day, it will be assigned an event class, i.e., event observation, for that day. One benefit of averaging the embeddings is to reduce the memory consumption. For example, assuming we have 100 stocks, 300 clusters of news events and 10 news articles affiliated to different clusters for a stock in a day, we have to store \emph{$100\times C^{10}_{300}  \approx 4.2 \times 10^{20}$} probability values which need 520914879903 GB, making the memory overhead infeasible. Therefore, we have to average them to reduce the overhead. These probability values indicate how different news events (actually, the event classes) impact the stock price fluctuations.The probability values are the parameters of ECHMM, and we need to store them and update them on each iteration of training.

Due to the sparsity of events, the events may be missing for some stocks on some days. Though our framework supports prediction with missing events, to improve the prediction performance, we fill the missing values as the nearest event embedding in the previous days. Specifically, for a stock \emph{s}, if it has an event embedding on day $t$, but lacks event embeddings on day $t+1$ and $t+2$, we just set the event embeddings on day $t+1$ and $t+2$ as that on day $t$. The intuition behind is that the impact of an event probably lasts for more than one day. Note that we also try other implementations in this work, involving without filling and partly filling, and the evaluation results will be shown in Sec.~\ref{results}.



\section{Extended Coupled Hidden Markov Model}
\label{model}

In this section, we will discuss how to use ECHMM for stock price movement prediction. We first introduce the notations used in the model, and then define the objective function. Finally, we will introduce the learning method for the model.


\subsection{Notations}

\begin{table}[!ht]
  \centering
   \caption{Notations and their meanings}
   \scriptsize
  \begin{tabular}{ll}  
  \hline
Notations    &Meanings\\ \hline  
$S$   &The number of stocks \\
$T$ &The number of time intervals\\
$H$ &The number of stock states\\
$z^{h}_{t, s}$ & The probability of stock \emph{s} in state \emph{h} in time \emph{t}\\
$O_{t}$ &The observation set includes stock event and stock price in time \emph{t}\\
$C_{s}$   &The set of the correlated stocks of stock \emph{s}\\
$C_{s,k}$ &The $\mathit{k}$-th stock in $C_{s}$ \\
$M$ & The stock price\\
$E$ & The stock event\\
$m_{t, s}$ &The set of stock price observations for stock \emph{s} in time \emph{t} including \\&  open price, close price, high price and low price\\
$m_{t, s, i}$ & One stock price observation for stock \emph{s} in time \emph{t}, $m_{t, s, i}$$\in$$m_{t, s}$\\
$e_{t, s}$ &The set of stock event observations for stock \emph{s} in time \emph{t}\\
$e_{t, s, i}$ &One stock event observation for stock \emph{s} in time \emph{t}, $e_{t, s, i}$$\in$$e_{t, s}$\\
$\Theta$ &The set of parameters of ECHMM \\
$\mathbf{\pi}$ &The matrix of the initial probability for stocks in each state\\
$\pi^{h}_{s}$ &The initial probability of stock \emph{s} in state \emph{h}\\
$A_{s}$ &The matrix of the state transition probability for stock \emph{s}\\
$g^{h}_{s}(\cdot)$ &The probability density of stock price for stock \emph{s} in time \emph{t}\\
$l^{h}_{s}(\cdot)$ &The probability density of stock event for stock \emph{s} in time \emph{t}\\
$q^{R_{i}, h}_{t, s}$ &The probability of a stock \emph{s} in state \emph{h} in time \emph{t} given that\\
 & its correlated stocks $C_{s}$ are in state $R_{i}=(r_{i1}, r_{i2},..., r_{i|N_l|})$ in time \emph{t-1} \\ \hline
\end{tabular}
\label{notation}
\end{table}

The notations used in our model and their corresponding meanings are shown in Table~\ref{notation}. The initial probability matrix, state transition probability matrix and emission matrix are three essential parameter matrices of ECHMM and we use $\bold{\Theta}$ to represent them. We use $\bold{\pi}$ to represent the initial probability matrix for stocks in each state and $\bold{A}$ to represent the state transition probability matrix. Let \emph{$g^{h}_{s}(\cdot)$} denote the probability density function of stock price for stock \emph{s} in state \emph{h} and assume it follows Gaussian distribution. Similarly, for the stock event, we use \emph{$l^{h}_{s}(\cdot)$} to denote the probability density function of stock event and assume it follows multinomial distribution. The stock event for stock \emph{s} in time \emph{t} is represented as \emph{$e_{t, s}$}. We use $\bold{O}$ to represent the set of the observations involving both the stock event observations and stock price observations. We also use \emph{$z^{h}_{t, s}$} to represent the probability of a stock \emph{s} being in state \emph{h} in time \emph{t}. Let \emph{$q^{R_{i}, h}_{t, s}$} represent the probability of a stock \emph{s} in state \emph{h} in time \emph{t} given that its correlated stocks $C_{s}$ are in state $R_{i}=(r_{i1}, r_{i2},..., r_{i|N_l|})$ in time \emph{t-1}.

\subsection{The Proposed Approach}
\label{papproach}
In this subsection, we will show how to obtain the log-likelihood of the observations and hidden variables, which can be described as

\begin{equation}\label{}
\emph{ $\log P(O ,Z \mid \Theta)= \log P(Z \mid  \Theta) + \log P(O \mid Z, \Theta)$} \\
\end{equation}

For the first term of formula (1), we can derive it as
\begin{equation}\label{}
\log P(Z \mid \Theta) =   \log P(Z_{1}) + \sum\limits^{T}_{t=2} \log P(Z_{t}\mid Z_{t-1})\\
\end{equation}

For the second term of formula (1), we can derive it according to the assumptions in Sec.~\ref{preliminary} and then get
\begin{equation}
    \begin{split}
\log P( O \mid Z, \Theta) =& \log P(M,E\mid Z, \Theta)=\log P(M\mid Z, \Theta)+\log P(E\mid Z, \Theta) \\
&= \sum\limits^{T}_{t=1} \log P(M_{t}\mid Z_{t}) + \sum\limits^{T}_{t=1} \log P(E_{t}\mid Z_{t}) \\
    \end{split}
\end{equation}

With the above analysis, we can rewrite the log likelihood of the hidden variables and observations of the ECHMM as follows:

\begin{equation}\label{}
 \begin{split}
\log P(O,Z\mid \Theta) &= \log P(Z_{1}) + \sum\limits^{T}_{t=2} \log P(Z_{t}\mid Z_{t-1})\\
&+ \sum\limits^{T}_{t=1} \log P(M_{t}\mid Z_{t}) + \sum\limits^{T}_{t=1} \log P(E_{t}\mid Z_{t})\\
\end{split}
\end{equation}

The first term of the formula (4) represents the initial probability of the stock states for the stock. The second term is the probability that stock state in time \emph{t-1} transits to a state in time \emph{t}. And the last two terms are the probability of the stock price observation and the probability of stock event observation conditioned on the stock states. Next, we will show how to compute these terms respectively.

The initial probability of the stock states of stock \emph{s} in the first time interval is

\begin{equation}
\begin{aligned}
\log P(Z_{1,s})= \sum\limits^{H}_{h=1} z^{h}_{1,s} \log \pi^{h}_{s}
\end{aligned}
\end{equation}

The log probability of stock state transiting from time \emph{t-1} to \emph{t} of stock \emph{s} can be derived as
\begin{equation}
\log P(Z_{t,s}|Z_{t-1,s})=\sum\limits^{H}_{h=1}\sum\limits^{H^{|C_{s}|}}_{i=1}(\prod\limits_{C_{s,j}\in C_{s}} z^{r_{ij}}_{t-1,C_{s,j}} z^{h}_{t, s}\log A_{s}(R_{i}, h))
\end{equation}

The second summation of formula (6) is over all the possible stock states \emph{$H^{|C_{s}|}$} of the set of correlation stocks, while the subsequent product is over terms on each of its individual correlation stock state given the state set \emph{($r_{i1}$,...,$r_{i|C_{s}|}$)}.

The probability of the stock price observation of stock \emph{s} given the stock states can be represented as
\begin{equation}
\log P(M_{t,s}\mid Z_{t,s})=\sum\limits^{H}_{h=1} z^{h}_{t, s}(\sum\limits_{m_{t, s, i}\in m_{t, s}} \log (g^{h}_{s}(m_{t, s, i})))
\end{equation}

The probability of the stock event observation of \emph{s} given the stock states can be represented as
\begin{equation}
\log P(E_{t,s}\mid Z_{t,s})=\sum\limits^{H}_{h=1} z^{h}_{t, s}(\sum\limits_{e_{t, s, i}\in e_{t, s}} \log (l^{h}_{s}(e_{t, s, i})))
\end{equation}

In this paper, the complete log-likelihood is critical since it can capture the characteristics of the stock direction changes during a period of time and can be trained by EM algorithm.

\subsection{Parameter Inference}


In this section, we apply the EM algorithm for finding the maximum-likelihood estimation of the parameters of ECHMM given a set of observed feature vectors and then we will get the complete log-likelihood of observations. This algorithm is also known as the Baum-Welch algorithm. Given the distribution of observations and the state transition matrix \emph{$A_{s}$}, it is possible to estimate the stock states based on the observations. Meanwhile, given the stock states of the stocks, we can estimate the parameters in the model.

EM algorithm has two steps: expectation step and maximization step. EM algorithm conducts the following two steps repeatedly until convergence:

1) Calculate the expected value of the log likelihood function, with respect to the conditional distribution of \emph{Z} given \emph{O} under the current estimate of the parameters \emph{$\Theta^{n}$}:
\begin{equation}
Q(\Theta,\Theta^{n+1}) = E_{Z}[\log[P(O, Z;\Theta)|O,\Theta^{n}]
\end{equation}

2) Set \emph{$\Theta^{n+1}= \mathop{\arg\max}_{\Theta} Q(\Theta,\Theta^{n+1})$}

The rest of this section will focus on deriving the necessary update steps to run this algorithm in our model.

\emph{\textbf{E-step.}} In E-step, we will calculate the expected value of the complete log-likelihood. We also get the transition probabilities given the parameters of \emph{$\Theta$} which include stock price information and stock event information, distribution parameters of the stock price and the state transition probability matrix.

According to Section~\ref{papproach}, we have already obtained \emph{$\log P(O\mid Z, \Theta)$}, and here we derive the expected complete log-likelihood of stock \emph{s} as follows:

\begin{equation}
    \begin{split}
    Q(\Theta,\Theta^{n+1}) &= E_{Z}[\log[P(O,Z;\Theta)|O,\Theta^{n}]  \\
    &=\sum\limits^{H}_{h=1} \log [P(O,Z|\Theta)]P(O,Z|\Theta^{n})\\
    &= \sum\limits^{H}_{h=1} \sum\limits^{T}_{t=1} z^{h}_{t, s} (\sum\limits_{e_{t, s, i }\in e_{t, s}}
         \log (l^{h}_{s}(e_{t, s, i})) + \sum\limits_{m_{t, s, i}\in m_{t, s}} \log (g^{h}_{s}(m_{t, s, i}))) \\
    & + \sum\limits^{T}_{t=2}\sum\limits^{H}_{h=1} \sum\limits^{H^{|C_{s}|}}_{i=1} q^{R_{i}, h}_{t, s}  (\log A_{s} (R_{i},h))+\sum\limits^{H}_{h=1} z^{h}_{1,s} \log \pi^{h}_{s}\\
    \end{split}
\end{equation}


Then we compute $z^{h}_{t, s}$ and $q^{R_{i}, h}_{t, s}$. \emph{$q^{R_{i}, h}_{t, s}$} can be obtained based on \emph{$z^{h}_{t, s}$, $z^{h}_{t-1, s}$} and the correlated stocks of stock \emph{s}. Then we introduce how to compute the probability of stock \emph{s} in a state in time \emph{t}. Due to the high computational overhead of ECHMM, we will apply a sequential importance sampling based approach, that is, particle filtering~\cite{cheng2006particle}, to estimate the stock state probability since it can reduce the amount of calculation. The algorithm is shown in Algorithm 1.

 \begin{algorithm}[!ht]
\caption{Estimating Stock Direction Probability}
\hspace*{0.02in} {\bf Input:}
The set of parameters of the CHMM $\Theta$, Time intervals \emph{T} and Number of samples \emph{K}.\\
\hspace*{0.02in} {\bf Output:}
The stock direction probability
\begin{algorithmic}[1]
 \State Initialization: randomly sample \emph{K} samples;
\For{t = 1 : \emph{T}}
 \For{k=1 : \emph{K}}
  \For{s=1 : \emph{S}}
   \State generate the sample $y^{k}_{t, s}$ based on the sample $y^{k}_{t-1, s}$  and the \State state transition             probability of stock s and its correlation stocks
  \EndFor
        \State compute the weights $w^{t}_{k} = P(M_{t}, E_{t}\mid y^{k}_{t})$
 \EndFor
  \State generate K samples $\{y^{k}_{t, s}\} ^{K,S}_{k=1, s=1}$
       \State compute the normalized weights $\hat{w}^{t}_{k} = w^{t}_{k}/\sum\nolimits^{K}_{j=1}w^{t}_{j} $
       \For{k=1: \emph{K}}
         \State generate the sample $\{\hat{y}^{k}_{t, s}\}^{S}_{s=1}$ with the weights $\hat{w}^{t}_{k}$
       \EndFor
\EndFor
\State get all the samples $\{\hat{y}^{k}_{t, s}\}^{T, S, K}_{t=1, s=1, k=1}$
\For{s=1: \emph{S}}
 \For{t=1: \emph{T}}
   \State computing $z^{h}_{t, s} $ with the samples $\{\hat{y}^{k}_{t, s}\}^{K}_{k=1}$
 \EndFor
\EndFor
\State \Return result
\end{algorithmic}
\end{algorithm}

\emph{\textbf{M-step.}} After obtaining the expected complete log-likelihood in the E-step, we perform M-step to update the parameters \emph{$\Theta$}: the initial state probability \emph{$\pi$}, the two observation distribution function parameters and the transition probability matrix \emph{$A_{s}$}, by maximizing the expected complete log-likelihood.

\begin{equation}
 \begin{split}
 &\mathop{\arg\max}_{\Theta} E_{h}[\log[P(O, z;\Theta)|O,\Theta^{n}] = \\
 &\mathop{\arg\max}_{\Theta} \sum\limits^{H}_{h=1} \sum\limits^{T}_{t=1} z^{h}_{t, s} (\sum\limits_{e_{t, s, i}\in e_{t, s}}  \log (l^{h}_{s} (e_{t, s, i})) + \sum\limits_{m_{t, s, i}\in m_{t, s}} \log (g^{h}_{s}(m_{t, s, i}))) \\
    & + \sum\limits^{T}_{t=2}\sum\limits^{H}_{h=1}\sum\limits^{H^{|C_{s}|}}_{i=1} q^{R_{i}, h}_{t, s}q^{R_{i}, h}_{t, s} (\log A_{s}(R_{i},s)) +  \sum\limits^{H}_{h=1} z^{h}_{1,s} \log \pi^{h}_{s}
 \end{split}
\end{equation}

\subsection{Model training and predictions}

\begin{figure}
  \centerline{\includegraphics[width=\textwidth]{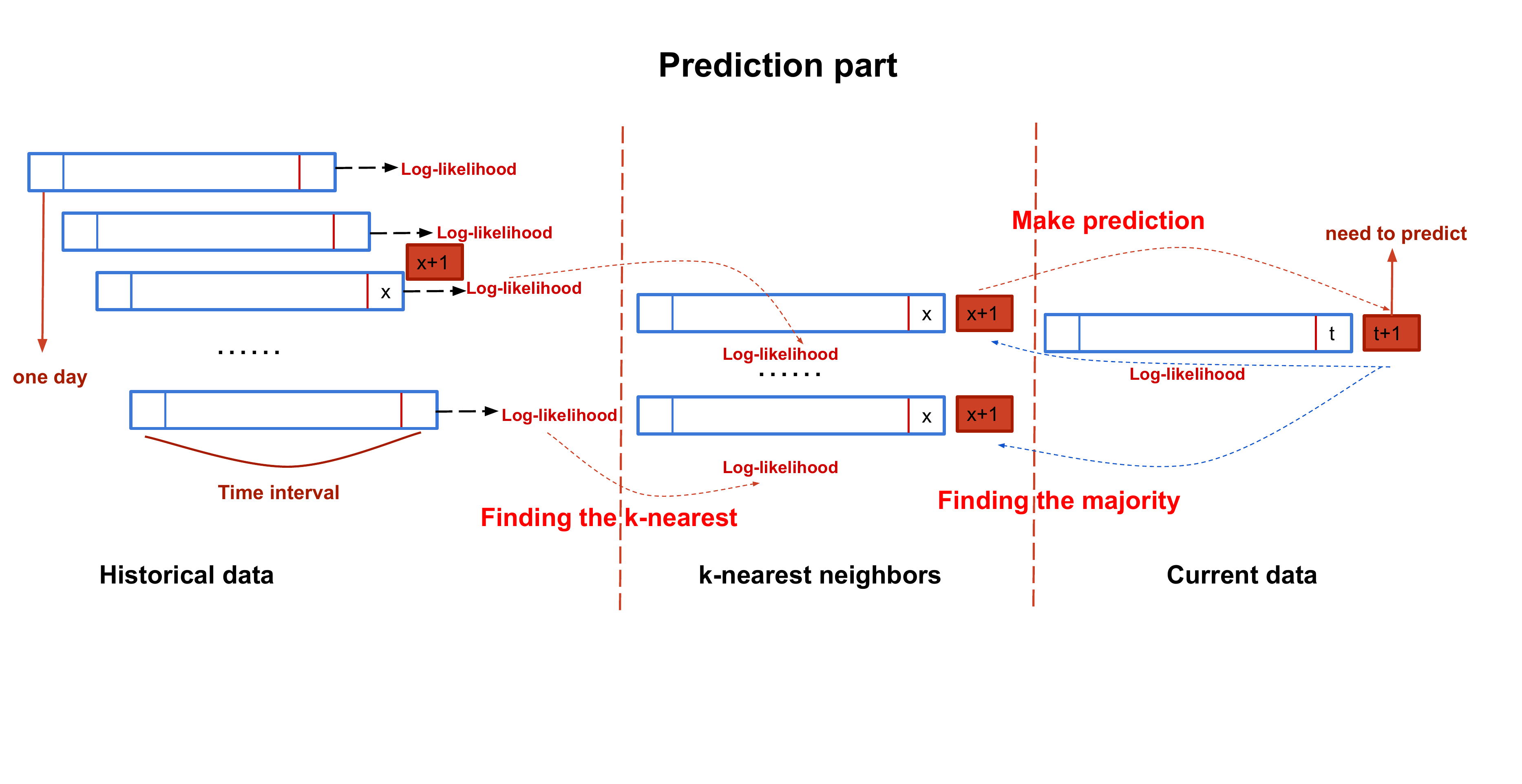}}
  \caption{Prediction method}
  \label{predict}
\end{figure}

With the trained ECHMM model, we next introduce how to perform stock movement prediction before each trading day starts. To involve the up-to-date stock fluctuation information in the training process, we use a dynamic training pool to train the parameters~\cite{zhang2004prediction}. Specifically, we set a time interval as a training pool, and train our model by computing the value of the expected complete log-likelihood within the training pool. At the end of a trading day, when the market closes, we will add the corresponding trading data of that day in the training pool, and meanwhile, drop the data of the first day in the training pool. We then update the model parameters with the new training pool. Fig.~\ref{predict} shows the prediction part.


We use the method of finding the $k$-nearest neighbors ($k$-NN) to make predictions, that is, an object is classified by a majority of its neighbors. Specifically, for the current day $t$, to predict the dirction of the next day, i.e., $h_{t+1}$, we will first compute the log-likelihood value $log_t$ of the current day, and then find  $k$ days that have $k$-nearest values from the historical data. Then, $h_{t+1}$ will be the diection most common among the directions of the next days of the $k$ neighbors. The intuition is that the price movement direction would be quite similar among the days that have close values of log-likelihood.

\label{}


\section{Experiments}
\label{experiment}
\subsection{Datasets}

We evaluate our proposed method with the China A-share stock market data during the period from Jan. 1, 2016 to Dec. 31, 2016. Due to the sparsity of data, we use the 100 stocks from the Chinese Stock Index (CSI) 100. We collect the corresponding event information and stock price information from Wind and our dataset is publicly available~\footnote{http://dwz.cn/stockdata}. We introduce the data source in detail as follows:

\emph{\textbf{Quantitative trading data.}} The quantitative trading data of stocks are collected from Wind~\footnote{http://www.wind.com.cn/}, a widely used financial information service provider in China. The indices we selected are the stock prices (open price, close price, high price and low price) as we think they are good indicators for stock direction changes. We also collect the p-change values to calculate the stock correlation.

\emph{\textbf{Stock news data.}} 21,728 news articles including titles and publication time in 2016 are collected from The Wind in total, and each article has been assigned to the corresponding stock. These web news are originally aggregated by Wind from major financial news websites in China, such as  \\ http://finance.sina.com.cn and http://www.hexun.com. These titles are then processed to extract events as described in Sec.~\ref{framework3}.

In the experiments, the length of dynamic training pool is set as 10 days, which is a relatively suitable time length to capture sufficient information and meanwhile keep sensitivity to the recent trend. We then divide all year data into a set of 10 day time intervals to train the model parameters and then obtain the log-likelihood value for the last day in each interval. We use the log-likelihood values in the first 10 months as the historical data from which to find the closest log-likelihood value and the intervals in the remaining 2 months as the testing samples. 45.9\% of the samples present upward trend, and 49.1\% present downward trend. The remaining 5\% keep still and we take the same price as positive signal. So the resulting labeled dataset is close to a balanced dataset.

\subsection{Comparision methods}

The following baselines and variations of our proposed model are implemented for comparison.

\emph{\textbf{SVM.}} We directly concatenate the stock price features (open price, close price, high price and low price) as well as stock event features as a linear vector, and then use them as the input of SVM for prediction.

\emph{\textbf{TeSIA.}} The tensor-based learning approach proposed in~\cite{li2015tensor} is the state-of-art baseline which utilizes multi-source information. Specifically, it uses a third-order tensor to model the firm-mode, event-mode, and sentiment-mode data. Note that they construct an independent tensor for every stock on each trading day, ignoring the correlations between stocks.

\emph{\textbf{CMT.}} CMT is a coupled matrix and tensor model proposed in~\cite{Zhang2018Improving}. It investigates the joint impacts of Web news and social media on the stock price movements via a coupled matrix and tensor factorization framework. Firstly, a tensor is
constructed by integrating various types of data to capture the intrinsic relations among events and sentiments. Due to the sparsity of the tensor, two auxiliary matrices, the stock-quantitative feature matrix and the stock correlation matrix, are incorporated to assist the tensor decomposition.

\emph{\textbf{ECHMM-NE.}} This is a variation of our proposed model. In order to verify that the stock news information is helpful for stock direction, we just apply the stock price information (stock price observations) on our model, without using the stock event observations.

\emph{\textbf{ECHMM-NC.}} This is another variation of our proposed model. In order to demonstrate that the stock correlation can facilitate stock prediction, we remove the stock correlation from our model and treat each stock's prediction as an independent task.

\emph{\textbf{ECHMM.}} This is the full implementation of our proposed ECHMM which not only uses stock correlations but also considers the joint effect of stock price information and stock event information.

Following the previous works~\cite{ding2015deep,ding2014using,Weng2017}, the standard measures of accuracy (ACC), Matthews Correlation Coefficient (MCC) and F1-score are used as metrics to evaluate each stock's price movement direction. Larger values of the metrics mean better classification performance. ACC is one of the most useful metrics for stock price movement prediction, however, it may fall short when two classes are of very different sizes. F1 and MCC are more useful than ACC when the class distribution is uneven. For F1, we have a range from 0 to 1: 0 when there are no true positives, undefined when there are only true negatives in the prediction, and 1 when there are neither false negatives nor false positives. Thus, F1 is a suitable measure for applications where true negatives don't matter. But in the application of stock prediction, the negatives (price going down) are also meaningful, which can be described by MCC, where true positives and true negatives are equally important. The MCC is in essence a correlation coefficient value between -1 and +1. A coefficient of +1 represents a perfect prediction, 0 no better than random prediction and -1 indicates total disagreement between prediction and observation. MCC is defined as

\begin{equation}
MCC = \frac{TP \times TN - FP \times FN}{\sqrt{(TP+FP)(TP+FN)(TN+FP)(TN+FN)}}
\end{equation}
where \emph{TP} and  \emph{TN} are the numbers of true positives and true negatives respectively, while \emph{FP} and  \emph{FN} denote the numbers of false positives and false negatives respectively in the confusion matrix. 

\subsection{Prediction results}
\label{results}

\begin{table}[!htbp]
   \centering
\caption{Prediction results of stock price movement direction}
  \begin{tabular}{llll}  %
    \hline
Method    &ACC  &F1 &MCC\\ \hline  
\textbf{SVM}   &52.21\% &0.5346 &0.0443\\
\textbf{TeSIA} &55.45\% &0.1818 &0.0143 \\
\textbf{CMT} &61.06\% &0.5275 &0.2102 \\
\textbf{ECHMM-NE} &60.48\% &0.7215 &0.0494\\
\textbf{ECHMM-NC} &59.43\% &0.7105 &0.0445\\
\textbf{ECHMM} &\textbf{62.70\%}  &\textbf{0.7390} &\textbf{0.0922}\\ \hline
\end{tabular}
 \label{predictionresults}
\end{table}

Table~\ref{predictionresults} shows the stock movement prediction performance for all the 60 stocks during the testing period. It can be observed that our proposed ECHMM achieves the best performance in terms of all the metrics. By contrast, SVM shows the worst performance, indicating that only using the linear combination of the features cannot capture the coupling effects, which are crucial for improving the performance. CMT achieves better performance in terms of ACC than other baselines, but still perform worse than ECHMM. The possible reason is that it neglects the time series information, which is crucial for stock prediction. TeSIA is only better than SVM in terms of ACC. Compared to ECHMM-NE and ECHMM-NC, ECHMM performs much better in all the metrics, we can observe the importance of stock news information and stock correlation information since ECHMM achieves better performance. We can also observe that TeSIA performs worse in MCC and F1 than other baselines, the possible reason is that TeSIA is not a time-series model, i.e., it uses only per day information, which may not be sufficient for prediction. In contrast, other methods use time-series historical data across a few days, allowing to explore more useful information and capture better moving trends. ECHMM presents a much higher result in terms of MCC than in terms of F1. The possible reason is that ECHMM can predict the negatives (price going down) quite well, since MCC cares more about negatives than F1 measure for binary classification. In addition, MCC has a larger range ([-1, +1]) than F1 ([0, 1]) may also account for part of the phenomenon.

\begin{table}[!h]
  \centering
\caption{Prediction results with varying lengths of historical data}
  \begin{tabular}{llll}  %
  \hline
Length of historical data   &ACC &F1 &MCC\\ \hline  
100 days  &62.10\% &0.7405 &0.0394\\
140 days &62.00\% &0.7384 &0.0519\\        
180 days  &62.22\% &0.7379 &0.0660 \\
220 days  &\textbf{62.70\%}  &\textbf{0.7390} &\textbf{0.0922}\\ \hline
\end{tabular}
  \label{lengthofhistory}
\end{table}

In order to observe whether the length of historical data will affect the results of prediction, we also conduct experiments with different lengths of historical data. Table~\ref{lengthofhistory} shows the stock movement direction prediction performance under various lengths of historical data. Obviously, it can be observed that the prediction performance improves as the length of historical data increases. The possible reason is that we are able to search for the closest log-likelihood in a larger range, and it is more likely to get a similar trend in the historical data.

\begin{table}[!h]
  \centering
\caption{Prediction results with varying amounts of stock events}
  \begin{tabular}{llll}  %
  \hline
Amounts of stock events   &ACC &F1 &MCC\\ \hline  
original events + fully filling events  &\textbf{62.70\%}  &\textbf{0.7390} &\textbf{0.0922}\\
only original events &60.00\% &0.7128 &0.0634\\        
80\% (original events + fully filling events)  &59.79\% &0.7090 &0.0763 \\
40\% (original events + fully filling events)  &58.87\% &0.6960 &0.0880\\ \hline
\end{tabular}
  \label{amountofevents}
\end{table}

The events extracted from the news are quite sparse. So we also conduct experiments to evaluate how event sparsity will influence the prediction performance, and the results are shown in Table~\ref{amountofevents}. It can be observed that the method that fully fills the missing events (original events + filled events) performs better than that without filling (i.e., only original events). We also evaluate with methods that work with only a part of events (80\% sampling and 40\% sampling), and they also perform worse than the fully filling method. We thus set the fully filling method as the default approach in this work. The method of filling the events is introduced in Sec.~\ref{framework3}.
\begin{figure}
\centering
  \begin{minipage}[c]{0.338\linewidth}
   \includegraphics[height=2cm,width=4.45cm]{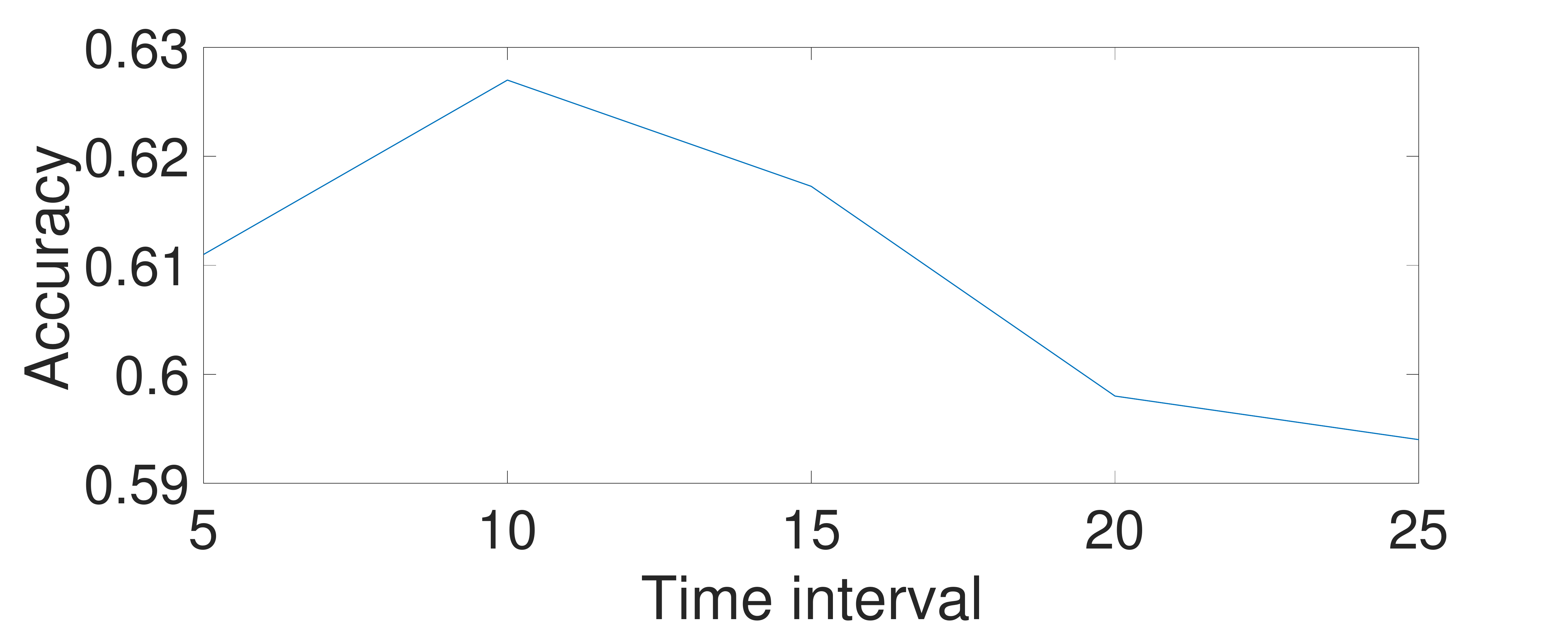}  
  \end{minipage}%
  \begin{minipage}[c]{0.338\linewidth}
   \includegraphics[height=2cm,width=4.45cm]{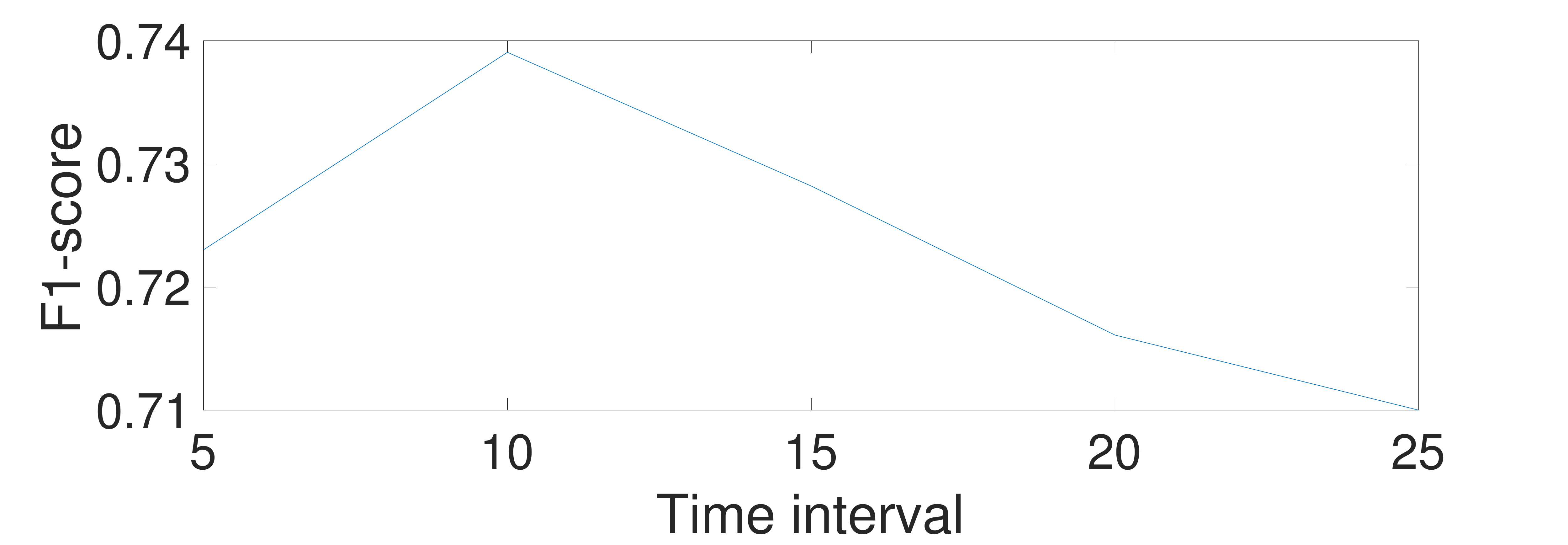}
   \end{minipage}%
  \begin{minipage}[c]{0.338\linewidth}
   \includegraphics[height=2cm,width=4.45cm]{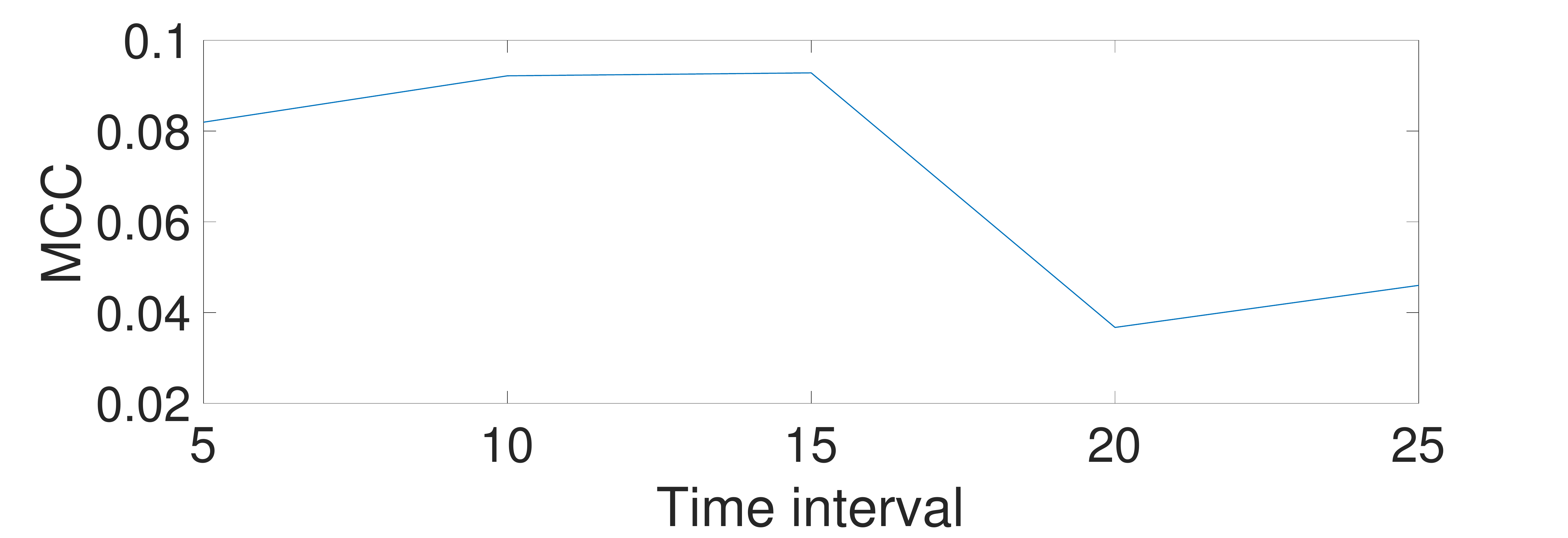}
   \end{minipage}%
  \caption{Prediction results with varying lengths of time interval}
  \label{timeinterval}
\end{figure}

We need to set the time interval of dynamic training pool to make predictions. So we conduct experiments on different lengths of such time intervals, and the results are shown in Fig.~\ref{timeinterval}. It can be observed that it achieves the best performance at 10 days. Thus, we set the length of dynamic training pool as 10 in this work.

\begin{figure}\centering
  \begin{minipage}[c]{0.339\linewidth}
   \includegraphics[height=2cm,width=4.5cm]{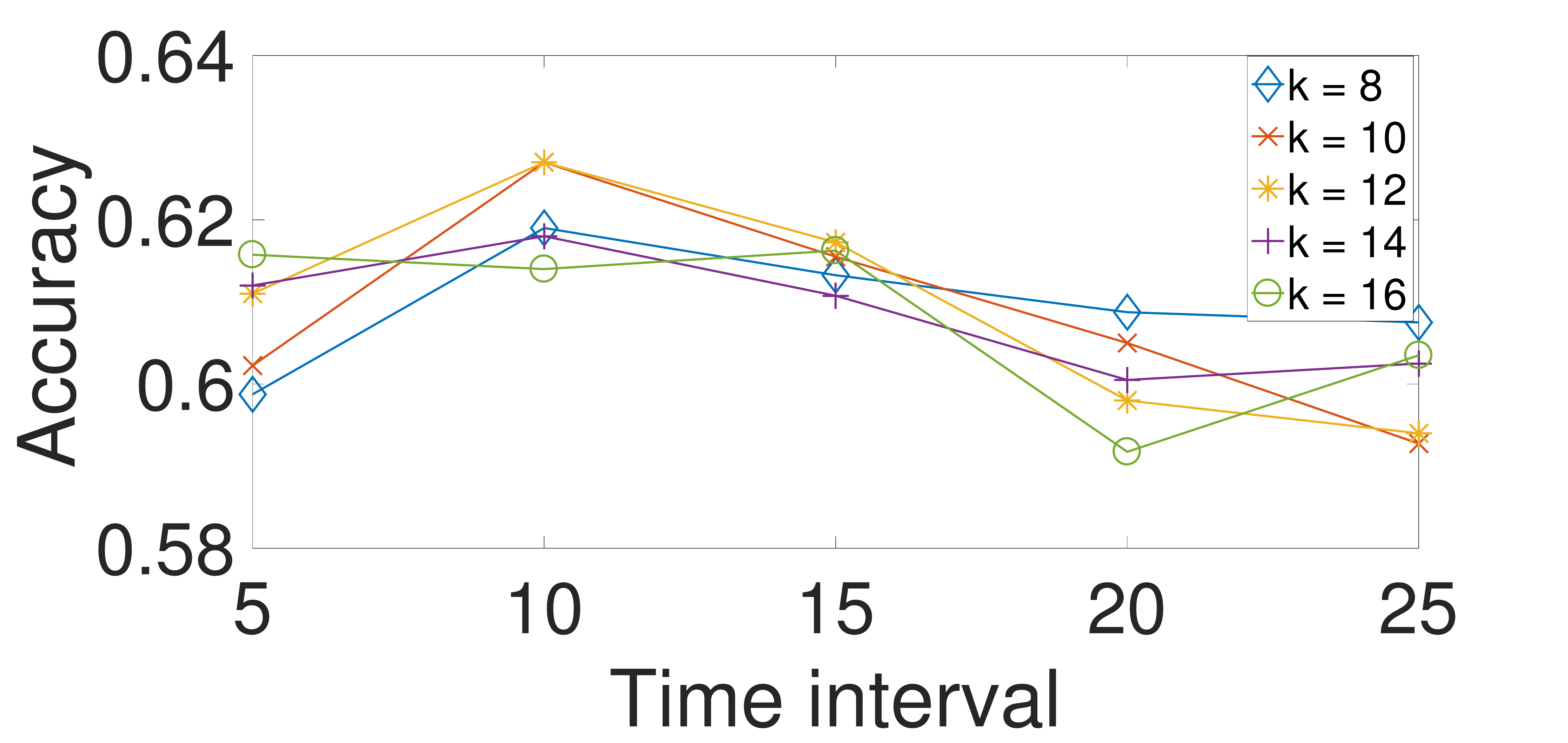}
   \end{minipage}%
  \begin{minipage}[c]{0.339\linewidth}
   \includegraphics[height=2cm,width=4.5cm]{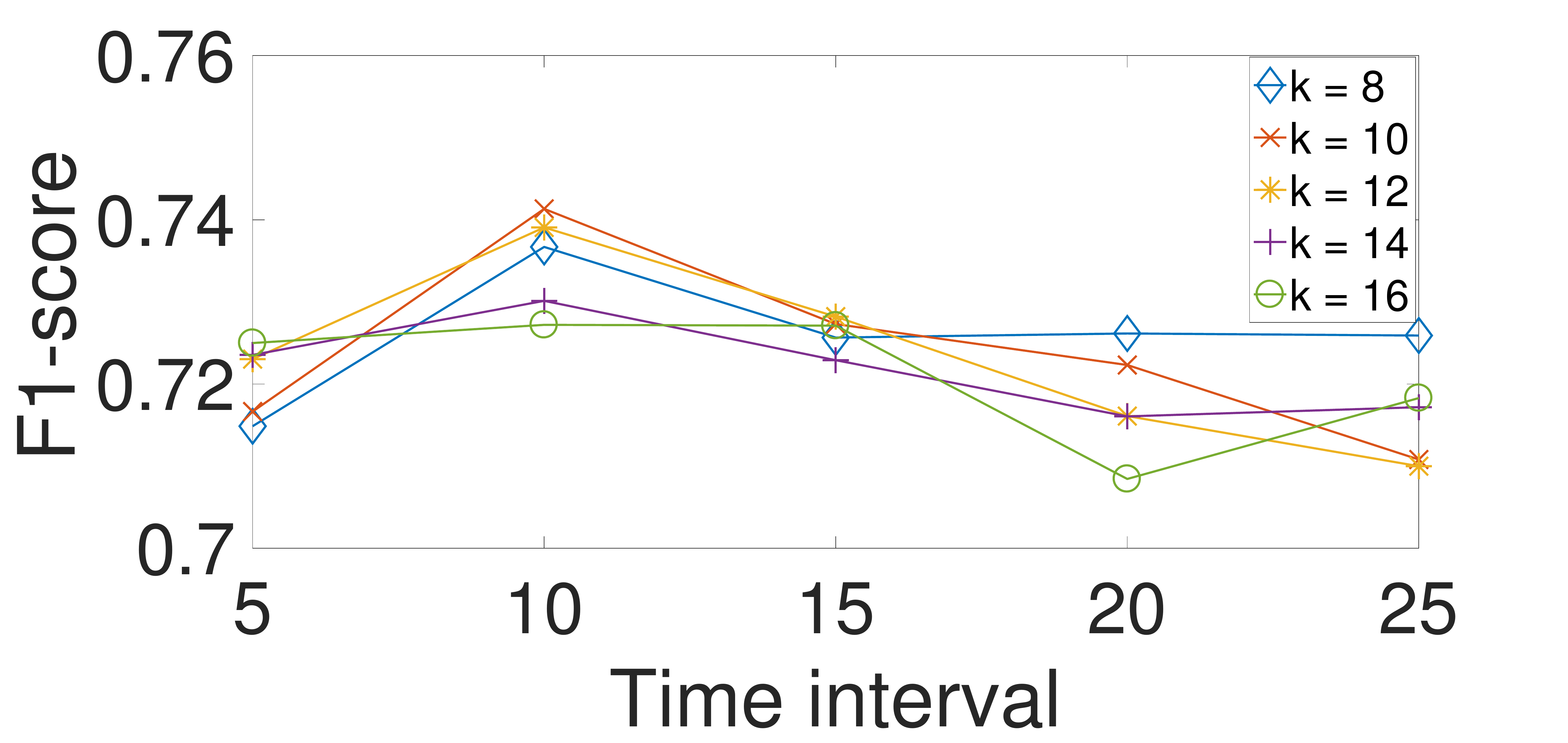}
   \end{minipage}%
  \begin{minipage}[c]{0.339\linewidth}
   \includegraphics[height=2cm,width=4.5cm]{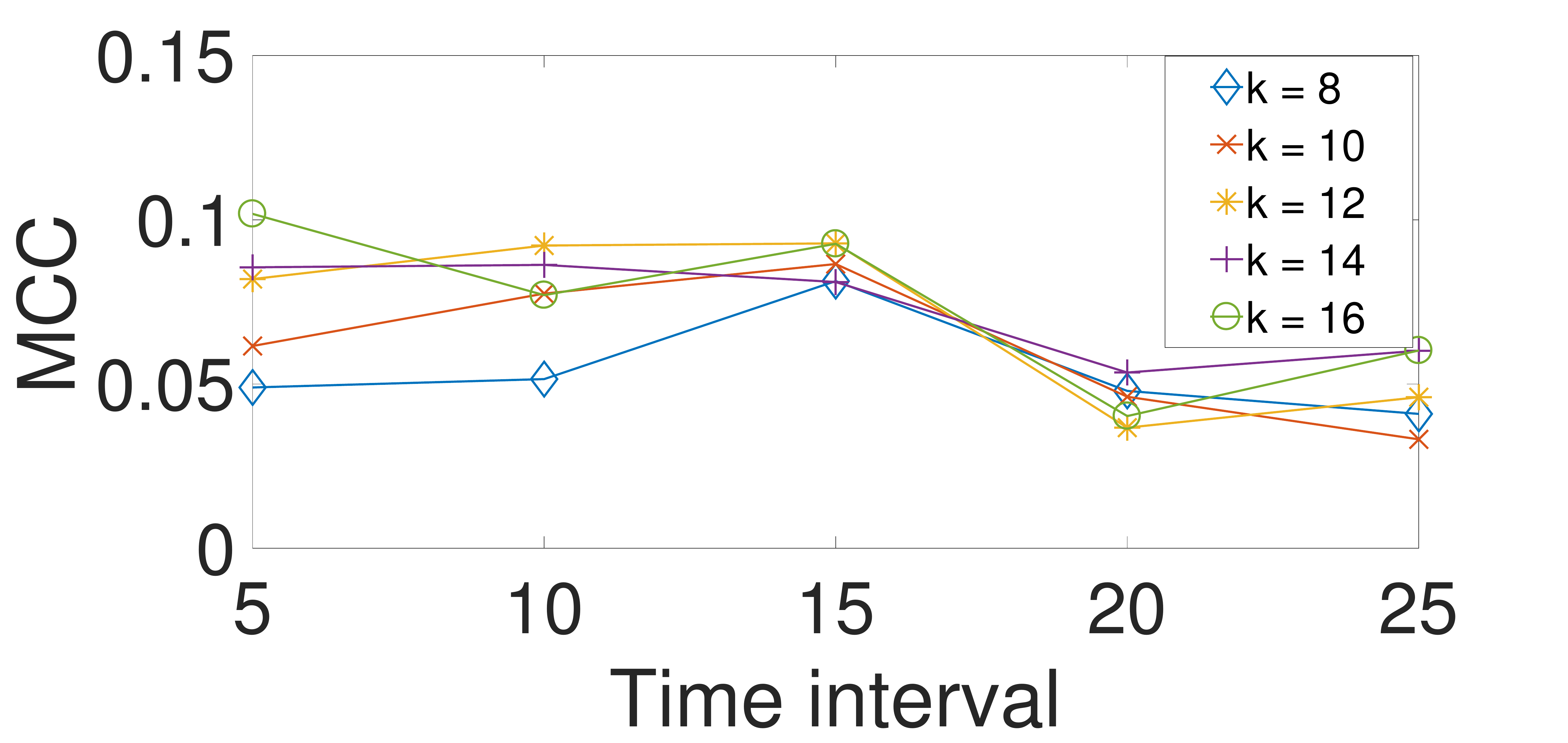}
   \end{minipage}%
  \caption{Prediction results with varying $k$ values}
  \label{k}
\end{figure}

We need to determine the value of $k$ for $k$-nearest neighbors ($k$-NN) method to make predictions. So we conduct experiments on different $k$ values of such time intervals, and the results are shown in Fig.~\ref{k}. It can be observed that it achieves the best performance at 12. Thus, we set the $k$ value of $k$-nearest neighbors ($k$-NN) method as 12 in this work.
\label{}
\section{Conclusions and future work}
\label{conclusion}

In this paper, we propose a time series information fusion framework, that is, the Extended Coupled Hidden Markov Model for stock prediction. Different from traditional methods that only consider the historical trading data or web news as features, our framework incorporates both of them to model their joint impacts. In addition, to alleviate the data sparsity problem, we utilize the stock correlations in the framework to facilitate each stock's prediction task. Evaluations on China A-share market in 2016 have demonstrated the effectiveness of our method.

Potential directions of future works include incorporating more data sources and applying more advanced NLP techniques to extract events. It would also be interesting to involve sentiment factors which have been proved to be able to drive stock fluctuations.

\label{}


\begin{acknowledgements}
This work has been supported by the State Key Development Program of Basic Research of China (No. 2013CB329604), the National Key Research and Development Program of China (No. 2016QY03D0605), the Natural Science Foundation of China (No. 61300014, 61672313), and NSF through grants IIS-1526499, IIS-1763325, and CNS-1626432, and DongGuan Innovative Research Team Program (No.201636000100038).
\end{acknowledgements}


\begin{thebibliography}{}

\providecommand{\natexlab}[1]{#1}

\bibitem{abdelaziz2015learning}
Abdelaziz AH, Zeiler S, Kolossa D (2015) Learning dynamic stream weights for
  coupled-hmm-based audio-visual speech recognition. IEEE/ACM Transactions on
  Audio, Speech and Language Processing (TASLP) 23(5):863--876

\bibitem{abushariah2010natural}
Abushariah MA, Ainon RN, Zainuddin R, Elshafei M, Khalifa OO (2010) Natural
  speaker-independent arabic speech recognition system based on hidden markov
  models using sphinx tools. In: Computer and Communication Engineering
  (ICCCE), 2010 International Conference on, IEEE, pp 1--6

\bibitem{akita2016deep}
Akita R, Yoshihara A, Matsubara T, Uehara K (2016) Deep learning for stock
  prediction using numerical and textual information. In: Computer and
  Information Science (ICIS), IEEE/ACIS 15th International Conference on, IEEE,
  pp 1--6

\bibitem{alsmadi2016information}
Alsmadi IM, Karabatis G, Aleroud A (2016) Information Fusion for Cyber-Security
  Analytics, vol 691. Springer

\bibitem{changmeasuring}
Chang CY, Zhang Y, Teng Z, Bozanic Z, Ke B (2016) Measuring the information
  content of financial news. In: Proceedings of the 26th International
  Conference on Computational Linguistics (COLING'16), pp 3216--3225

\bibitem{chen2015lstm}
Chen K, Zhou Y, Dai F (2015) A lstm-based method for stock returns prediction:
  A case study of china stock market. In: Big Data (Big Data), 2015 IEEE
  International Conference on, IEEE, pp 2823--2824

\bibitem{cheng2006particle}
Cheng P, Qiu Z, Ran B (2006) Particle filter based traffic state estimation
  using cell phone network data. In: Intelligent Transportation Systems
  Conference, 2006. ITSC'06. IEEE, IEEE, pp 1047--1052

\bibitem{Chiang2016}
Chiang Wc, Enke D, Wu T, Wang R (2016) {An adaptive stock index trading
  decision support system}. Expert Systems With Applications 59:195--207

\bibitem{Chong2017}
Chong E, Han C, Park FC (2017) {Deep learning networks for stock market
  analysis and prediction : Methodology , data representations , and case
  studies}. Expert Systems With Applications 83:187--205

\bibitem{cutler1988moves}
Cutler DM, Poterba JM, Summers LH (1989) What moves stock prices. J Portf Manag
  15:4--12

\bibitem{das2007yahoo}
Das SR, Chen MY (2007) Yahoo! for amazon: Sentiment extraction from small talk
  on the web. Management science 53(9):1375--1388

\bibitem{ding2014using}
Ding X, Zhang Y, Liu T, Duan J (2014) Using structured events to predict stock
  price movement: An empirical investigation. In: EMNLP, pp 1415--1425

\bibitem{ding2015deep}
Ding X, Zhang Y, Liu T, Duan J (2015) Deep learning for event-driven stock
  prediction. In: IJCAI, pp 2327--2333

\bibitem{Efficiency2016Effective}
Efficiency TI (2016) Effective and fast near duplicate detection via
  signature-based compression metrics 2016(8--13):1--12

\bibitem{elmezain2009hidden}
Elmezain M, Al-Hamadi A, Appenrodt J, Michaelis B (2009) A hidden markov
  model-based isolated and meaningful hand gesture recognition. International
  Journal of Electrical, Computer, and Systems Engineering 3(3):156--163

\bibitem{fama1965behavior}
Fama EF (1965) The behavior of stock-market prices. The journal of Business
  38(1):34--105

\bibitem{feldman2011stock}
Feldman R, Rosenfeld B, Bar-Haim R, Fresko M (2011) The stock sonar-sentiment
  analysis of stocks based on a hybrid approach. In: Twenty-Third IAAI
  Conference

\bibitem{Goumatianos2017}
Goumatianos N, Christou IT, Lindgren P, Prasad R (2017) An algorithmic
  framework for frequent intraday pattern recognition and exploitation in forex
  market. Knowledge and Information Systems

\bibitem{gupta2012stock}
Gupta A, Dhingra B (2012) Stock market prediction using hidden markov models.
  In: Engineering and Systems (SCES), 2012 Students Conference on, IEEE, pp
  1--4

\bibitem{Hai2015}
Hai T, Shirai K, Velcin J (2015) {Sentiment analysis on social media for stock
  movement prediction}. Expert Systems With Applications 42(24):9603--9611

\bibitem{hassan2005stock}
Hassan MR, Nath B (2005) Stock market forecasting using hidden markov model: a
  new approach. In: Intelligent Systems Design and Applications, 2005. ISDA'05.
  Proceedings. 5th International Conference on, IEEE, pp 192--196

\bibitem{hogenboom2011overview}
Hogenboom F, Frasincar F, Kaymak U, De~Jong F (2011) An overview of event
  extraction from text. In: Workshop on Detection, Representation, and
  Exploitation of Events in the Semantic Web (DeRiVE 2011) at Tenth
  International Semantic Web Conference (ISWC 2011), Citeseer, vol 779, pp
  48--57

\bibitem{jia2016investigation}
Jia H (2016) Investigation into the effectiveness of long short term memory
  networks for stock price prediction. arXiv preprint arXiv:160307893

\bibitem{kogan2009predicting}
Kogan S, Levin D, Routledge BR, Sagi JS, Smith NA (2009) Predicting risk from
  financial reports with regression. In: Proceedings of Human Language
  Technologies: The 2009 Annual Conference of the North American Chapter of the
  Association for Computational Linguistics, Association for Computational
  Linguistics, pp 272--280

\bibitem{kumar2017coupled}
Kumar P, Gauba H, Roy PP, Dogra DP (2017) Coupled hmm-based multi-sensor data
  fusion for sign language recognition. Pattern Recognition Letters 86:1--8

\bibitem{lavrenko2000mining}
Lavrenko V, Schmill M, Lawrie D, Ogilvie P, Jensen D, Allan J (2000) Mining of
  concurrent text and time series. In: KDD-2000 Workshop on Text Mining, pp
  37--44

\bibitem{li2015tensor}
Li Q, Jiang L, Li P, Chen H (2015) Tensor-based learning for predicting stock
  movements. In: AAAI, pp 1784--1790

\bibitem{li2014news}
Li X, Xie H, Chen L, Wang J, Deng X (2014) News impact on stock price return
  via sentiment analysis. Knowl-Based Syst 69:14--23

\bibitem{luss2015predicting}
Luss R, D'Aspremont A (2015) Predicting abnormal returns from news using text
  classification. Quantitative Finance 15(6):999--1012

\bibitem{mikolov2013distributed}
Mikolov T, Sutskever I, Chen K, Corrado GS, Dean J (2013) Distributed
  representations of words and phrases and their compositionality. In: Advances
  in neural information processing systems, pp 3111--3119

\bibitem{nefian2007coupled}
Nefian AV (2007) Coupled hidden markov model for audiovisual speech
  recognition. {U}S Patent 7,165,029

\bibitem{nguyen2013event}
Nguyen T, Phung D, Adams B, Venkatesh S (2013) Event extraction using behaviors
  of sentiment signals and burst structure in social media. Knowledge and
  information systems pp 1--26

\bibitem{nguyen2015topic}
Nguyen TH, Shirai K (2015) Topic modeling based sentiment analysis on social
  media for stock market prediction. In: Proceedings of the 53rd Annural
  Meeting of the Association for Computational Linguistics (ACL-15)

\bibitem{nickel2011using}
Nickel C, Busch C, Rangarajan S, M{\"o}bius M (2011) Using hidden markov models
  for accelerometer-based biometric gait recognition. In: Signal Processing and
  its Applications (CSPA), 2011 IEEE 7th International Colloquium on, IEEE, pp
  58--63

\bibitem{Oliveira2017}
Oliveira N, Cortez P, Areal N (2017) {The impact of microblogging data for
  stock market prediction : Using Twitter to predict returns , volatility ,
  trading volume and survey sentiment indices}. Expert Systems With
  Applications 73:125--144

\bibitem{park2009forecasting}
Park SH, Lee JH, Song JW, Park TS (2009) Forecasting change directions for
  financial time series using hidden markov model. Rough Sets and Knowledge
  Technology pp 184--191

\bibitem{patra2009computationally}
Patra JC, Thanh NC, Meher PK (2009) Computationally efficient flann-based
  intelligent stock price prediction system. In: Neural Networks, 2009. IJCNN
  2009. International Joint Conference on, IEEE, pp 2431--2438

\bibitem{peng2015leverage}
Peng Y, Jiang H (2015) Leverage financial news to predict stock price movements
  using word embeddings and deep neural networks. arXiv preprint
  arXiv:150607220

\bibitem{Poh2000}
Poh KL (2000) An intelligent decision support system for investment analysis.
  Knowledge and Information Systems 2(3):340--358

\bibitem{rabiner1989tutorial}
Rabiner LR (1989) A tutorial on hidden markov models and selected applications
  in speech recognition. Proceedings of the IEEE 77(2):257--286

\bibitem{roman1996backpropagation}
Roman J, Jameel A (1996) Backpropagation and recurrent neural networks in
  financial analysis of multiple stock market returns. In: System Sciences,
  1996., Proceedings of the Twenty-Ninth Hawaii International Conference on,,
  IEEE, vol~2, pp 454--460

\bibitem{saad1998comparative}
Saad EW, Prokhorov DV, Wunsch DC (1998) Comparative study of stock trend
  prediction using time delay, recurrent and probabilistic neural networks.
  IEEE Transactions on neural networks 9(6):1456--1470

\bibitem{schumaker2009quantitative}
Schumaker RP, Chen H (2009{\natexlab{a}}) A quantitative stock prediction
  system based on financial news. Information Processing \& Management
  45(5):571--583

\bibitem{schumaker2009textual}
Schumaker RP, Chen H (2009{\natexlab{b}}) Textual analysis of stock market
  prediction using breaking financial news: The azfin text system. ACM
  Transactions on Information Systems (TOIS) 27(2):12

\bibitem{si2013exploiting}
Si J, Mukherjee A, Liu B, Li Q, Li H, Deng X (2013) Exploiting topic based
  twitter sentiment for stock prediction. ACL (2) 2013:24--29

\bibitem{si2014exploiting}
Si J, Mukherjee A, Liu B, Pan SJ, Li Q, Li H (2014) Exploiting social relations
  and sentiment for stock prediction. In: The Conference on Empirical Methods
  on Natural Language Processing (EMNLP-14), vol~14, pp 1139--1145

\bibitem{tetlock2007giving}
Tetlock PC (2007) Giving content to investor sentiment: The role of media in
  the stock market. The Journal of Finance 62(3):1139--1168

\bibitem{tetlock2008more}
Tetlock PC, SAAR-TSECHANSKY M, Macskassy S (2008) More than words: Quantifying
  language to measure firms' fundamentals. J Finance 63(3):1437--1467

\bibitem{MMRate}
Wang S, Hu X, Yu P, Li Z (2014) Mmrate: Inferring multi-aspect diffusion
  networks with multi-pattern cascades. In: Proceedings of the ACM SIGKDD
  International Conference on Knowledge Discovery and Data Mining, pp
  1246--1255

\bibitem{wang2016enhancing}
Wang S, Li F, Stenneth L, Philip SY (2016) Enhancing traffic congestion
  estimation with social media by coupled hidden markov model. In: Joint
  European Conference on Machine Learning and Knowledge Discovery in Databases,
  Springer, pp 247--264

\bibitem{wang2014semiparametric}
Wang WY, Hua Z (2014) A semiparametric gaussian copula regression model for
  predicting financial risks from earnings calls. In: ACL (1), pp 1155--1165

\bibitem{Weng2017}
Weng B, Ahmed MA, Megahed FM (2017) {Stock market one-day ahead movement
  prediction using disparate data sources} 79:153--163

\bibitem{xie2013semantic}
Xie B, Passonneau RJ, Wu L, Creamer GG (2013) Semantic frames to predict stock
  price movement. In: Proceedings of the 51st Annual Meeting of the Association
  for Computational Linguistics, pp 873--883

\bibitem{yoshihara2014predicting}
Yoshihara A, Fujikawa K, Seki K, Uehara K (2014) Predicting stock market trends
  by recurrent deep neural networks. In: Pacific Rim International Conference
  on Artificial Intelligence, Springer, pp 759--769

\bibitem{Zhang2014Discovering}
Zhang J, Shi X, Kong X, Shuai HH, Yu PS (2014) Discovering organizational
  correlations from twitter. In: IEEE International Conference on Data Mining
  Workshop, pp 243--250

\bibitem{zhang2011predicting}
Zhang X, Fuehres H, Gloor PA (2011) Predicting stock market indicators through
  twitter ?{I} hope it is not as bad as {I} fear. Proc-Soc Behav Sci 26:55--62

\bibitem{ZHANG2017}
Zhang X, Shi J, Wang D, Fang B (2017) Exploiting investors social network for
  stock prediction in china's market. Journal of Computational Science

\bibitem{Zhang2018Improving}
Zhang X, Zhang Y, Wang S, Yao Y, Fang B, Yu PS (2018) Improving stock market
  prediction via heterogeneous information fusion. Knowledge-Based Systems

\bibitem{zhang2004prediction}
Zhang Y (2004) Prediction of financial time series with hidden markov models.
  Master's thesis, School of Computing Science, Simon Fraser University

\bibitem{Zhu:2016:CFL:3016387.3016423}
Zhu W, Lan C, Xing J, Zeng W, Li Y, Shen L, Xie X (2016) Co-occurrence feature
  learning for skeleton based action recognition using regularized deep lstm
  networks. In: Proceedings of the Thirtieth AAAI Conference on Artificial
  Intelligence, AAAI'16, pp 3697--3703




%
%
%
%
%
%
%
%
%
%
%
%
%
%
%
%
%
%
%
%
%
%
%
%
%
%
%
%
%
%
%
%
%
%
%
%
%
%
%
%
%
%
%
%
%
%
%
%
%
%
%
%
%
%
%
%
%
%
%
%
%
%
%

\end{thebibliography}


\end{document}